\documentclass[reprint,amsmath,amssymb,aps,superscriptaddress]{revtex4-1}
\usepackage{graphicx}
\usepackage{dcolumn}
\usepackage{bm}
\usepackage{hyperref}
\begin{document}

\title{Sign reversal of anomalous Hall conductivity and magnetoresistance in cubic non-collinear antiferromagnet Mn$_3$Pt thin films}

\author{Joynarayan Mukherjee}
\email[Corresponding author: ]{mjoynarayan@tifrh.res.in}
\affiliation{Tata Institute of Fundamental Research, Hyderabad-500107, India}
\author{T. S. Suraj}
\affiliation{Low temperature Physics lab, Department of Physics, IIT Madras, Chennai-600036, India}

\author{Himalaya Basumatary}
\affiliation{Advanced magnetic group, DMRL, Hyderabad, India}
\author{K. Sethupathi}
\affiliation{Low temperature Physics lab, Department of Physics, IIT Madras, Chennai-600036, India}

\author{Karthik V. Raman}
\email[Corresponding author: ]{kvraman@tifrh.res.in}
\affiliation{Tata Institute of Fundamental Research, Hyderabad-500107, India}


\begin{abstract}
The two dimensional kagome spin lattice structure of Mn atoms in the family of Mn$_3$X non-collinear antiferromagnets are providing substantial excitement in the exploration of Berry curvature physics and the associated non-trivial magnetotransport responses. Much of these studies are performed in the hexagonal systems, mainly Mn$_3$Sn and Mn$_3$Ge, with the kagome planes having their normal along the [001] direction. In this manuscript, we report our study in the cubic Mn$_3$Pt thin films with their kagome planes normal to the [111] crystal axis. Our studies reveal a hole conduction dominant Hall response with a non-monotonic temperature dependence of anomalous Hall conductivity (AHC), increasing from 9 $\Omega^{-1}$cm$^{-1}$ at room temperature to 29 $\Omega^{-1}$cm$^{-1}$ at 100 K, followed by a drop and unexpected sign-reversal at lower temperatures. Similar sign reversal is also observed in magnetoresistance measurements. We attribute this sign reversal to the transition from a Berry curvature dominated AHC at high temperature to a weak canted ferromagnetic AHC response at lower temperature, below 70 K, caused by the reorientation of Mn moments out of the kagome plane. Our above results in thin films of  Mn$_3$Pt make advances in their integration with room temperature antiferromagnetic spintronics.

\end{abstract}
\maketitle
\section{Introduction}
The hexa-triangular (kagome) spin lattice arrangement in non-collinear antiferromagnets (NCAFMs) are attracting significant attention due to their non-trivial electronic band structure leading to the exploration of novel topological properties \cite{Kiyohara2016_AHE_Mn3Ge,Higo2018_MOKE_Mn3Sn,Lin2018_Flatbands_Fe3Sn2,Yin2018_flatbands_Fe3Sn2,Chen2014_AHE_Theory_Mcdonald,Kuebler2014_AHE_theory_Felser}. Specifically, the kagome spin structure gives rise to non-vanishing momentum space berry
curvature \cite{Chen2014_AHE_Theory_Mcdonald,Kuebler2014_AHE_theory_Felser} supporting novel phenomena such
as anomalous Hall effect (AHE) \cite{Chen2014_AHE_Theory_Mcdonald,Kuebler2014_AHE_theory_Felser,Nakatsuji2015_AHE_Mn3Sn}, spin Hall effect \cite{Zhang2016_SHE_Mn3Ir}, magnetic spin Hall effect \cite{Kimata2019_MSHE},
topological Hall effect \cite{Rout2019_THE_Mn3Sn}, magneto-optical Kerr effect \cite{Higo2018_MOKE_Mn3Sn}
and anomalous Nernst effect \cite{Li2017_ANE_Righileduc_Mn3Sn}. Furthermore, they are predicted to host quantum anomalous Hall state \cite{Xu2015_Kagome_QAHE_CsLiMnF,Kobayashi2019_Kagome_QAHE}, Weyl fermion excitations  \cite{Nakatsuji2015_AHE_Mn3Sn} and support non-dispersive flat band near the Fermi level with topological semimetal response \cite{Lin2018_Flatbands_Fe3Sn2,Yin2018_flatbands_Fe3Sn2}. All these properties make NCAFMs a potential materials candidate in the field of spintronics and, also, topo-electronics. 

\begin{figure*}
\includegraphics[width=16cm]{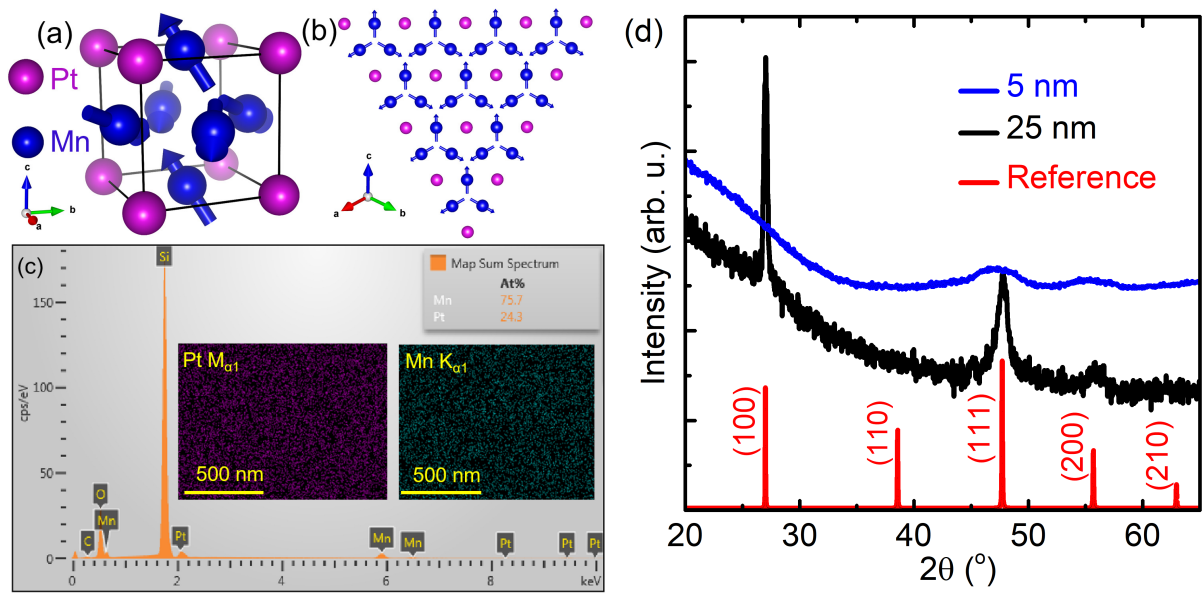}\caption{(a) Crystal and magnetic structure of cubic Mn$_3$Pt with
space group $Pm\bar{3}m$. (b) (111) plane of Mn$_3$Pt
showing the kagome structure of Mn moments with all-in all-out spin configuration. (c) EDS spectra of 25 nm Mn$_3$Pt thin film. Insets show elemental mapping of Mn and Pt. (d) X-ray diffraction
pattern of 5 and 25 nm Mn$_3$Pt thin films grown on
Si/SiO$_2$ substrate. Red line is the simulated
XRD pattern of Mn$_3$Pt considering Co $K_{\alpha}$ source. The crystal structure and the XRD pattern were generated using VESTA software \cite{VESTA_Momma2011}. \label{fig:(a)-X-ray-diffraction}}
\end{figure*}

NCAFM ordering is observed in various materials such as in perovskite oxides ((Y, Lu)MnO$_3$, SmFeO$_3$) \cite{Oh2016_RmnO3,Hajiri2019_SmFeO3}, nitrides (Mn$_3$GaN) \cite{Hajiri2019_Mn3GaN}, Mn$_5$Si$_3$ \cite{Suergers2014_Mn5Si3_THE} and in Mn$_3$X (X = Ir, Pt, Sb, Sn, Ge, Ga) \cite{Zhang2017_Mn3X_DFT_structure}. Among these, the Mn$_3$X family of materials are a very promising choice due to their ease-in-growth achievable using co-evaporation process with precise control of the stoichiometry. Here, the triangular or inverse triangular arrangement of Mn spin moments in the kagome plane show strong electronic correlations with spin-orbit effects leading to non-trivial band structure. Mn$_3$X compounds crystallize in simple cubic (X=Ir, Pt, Sb) and in hexagonal (X=Sn, Ge, Ga) structure with their kagome planes along the (111) and (001) crystal plane, respectively. These structures macroscopically give rise to zero magnetic moment due to the cancellation of effective moments from the three sub-lattices (as shown in fig. \ref{fig:(a)-X-ray-diffraction}(a) and (b)) \cite{Zhang2017_Mn3X_DFT_structure,Liu2018_Mn3Pt_epitaxial_AHE}. Experimentally, Mn$_3$X compounds are observed to undergo temperature driven phase transitions between collinear AFM, non-collinear AFM and/or canted ferromagnetism state depending on the composition, microstructure and growth procedure \cite{Sung2018_Mn3Sn_SC_AHE_phase,Ji2006_MnPt_TF_Phasediagram}. In the collinear AFM state, the Berry curvature becomes zero due to the symmetry \cite{Nayak2016_AHE_Mn3Ge}, while in the non-collinear phase, this special spin arrangement of Mn moments support finite Berry curvature \cite{Higo2018_MOKE_Mn3Sn} leading to the observation of anomalous Hall conductivity (AHC). This observation is of particular interest since AHE allows us to correlate the magnetic and electrical transport properties of the material \cite{Chen2014_AHE_Theory_Mcdonald,Kuebler2014_AHE_theory_Felser}. In literature, large AHC has been reported in Mn$_3$Sn \cite{Nakatsuji2015_AHE_Mn3Sn} and Mn$_3$Ge \cite{Nayak2016_AHE_Mn3Ge} single crystals showing few tens of $\Omega^{-1}$cm$^{-1}$ at 300 K to few hundreds of $\Omega^{-1}$cm$^{-1}$ at low temperature that are comparable
to reported values in ferromagnetic metals ($\sim$ 500 $\Omega^{-1}$cm$^{-1}$) \cite{Miyasato2007_AHE_Fe_Co_Ni,Yao2004_AHE_Fe_DFT}. In both Mn$_3$Sn and Mn$_3$Ge, AHC vanishes or becomes negligible when the magnetic field is applied parallel to the mirror symmetry plane of the hexagonal lattice suggesting a strong directionality of the AHC response. AHE is also observed in thin films of Mn$_3$Sn and Mn$_3$Ge grown on suitable substrates \cite{You2018_Mn3Sn_Epi_AHE,qin2020_AHE_Mn3Ge}. In epitaxially grown Mn$_3$Sn thin films, AHC was found to be 21 $\Omega^{-1}$cm$^{-1}$ at 300 K \cite{Taylor2020_Mn3SnTF_AHE_THE} while a sign change in AHC was reported in polycrystalline Mn$_3$Sn films below 200 K \cite{Ikeda2018_Mn3Sn_Poly_AHE,PhysRevB.99.184425}.

\par In this paper, we investigate the AHE and magnetoresistance (MR) simultaneously and correlate these properties in an under-studied cubic structure of Mn$_3$Pt (space group $Pm\bar{3}m$) films with a bulk lattice parameter of 3.833 \AA. The (111) kagome plane of cubic Mn$_3$Pt has an all-in all-out spin configuration \cite{Liu2018_Mn3Pt_epitaxial_AHE,Zhang2017_Mn3X_DFT_structure}, as shown in figure \ref{fig:(a)-X-ray-diffraction}b. Here, the Mn - Mn spacing in the kagome plane is significantly smaller at 2.710 {\AA} than in the case of hexagonal structure Mn$_3$Sn (2.925 {\AA}). As a result, Mn$_3$Pt may support interesting and distinct electronic spin-correlations. The N\'eel temperature of Mn$_3$Pt is 475 K, corresponding to the paramagnetic to AFM transition. At 365 K, Mn$_3$Pt undergoes a first order phase transition from collinear AFM to NCAFM \cite{Kren1967_PhaseTransition_Mn3Pt,Liu2018_Mn3Pt_epitaxial_AHE}. Recently, strain tunable AHE was observed in epitaxial (100)-Mn$_3$Pt thin films, grown on ferroelectric BaTiO$_3$(100) substrate, showing a maximum AHC of 98$\pm$13 $\Omega^{-1}$cm$^{-1}$ at 10 K \cite{Liu2018_Mn3Pt_epitaxial_AHE}.

\section{Thin film preparation and structural characterization}
We prepared thin films of Mn$_3$Pt with varying thickness (5 - 25 nm) on Si/SiO$_2$ substrate using a co- d.c. sputtering of Mn ($>99.9$ purity) and Pt ($>99.9$ purity) targets in our custom-designed sputtering chamber with a base pressure of $5\times10^{-9}$ mbar. For the thin film growth, substrate temperature and Ar partial pressure were fixed at 450 \textsuperscript{o}C and $5\times10^{-3}$ mbar, respectively. A 2 nm thin seed layer of Ta was grown at room temperature to prevent interfacial oxidation of the films from the SiO$_2$ substrate. Mn$_3$Pt films were subsequently annealed for 15 min at 450 \textsuperscript{o}C in a vacuum level of $\sim1\times10^{-8}$ mbar, followed by a 3.5 nm capping layer of Ta deposited at room temperature. Energy dispersive X-ray spectroscopy (EDS) analysis with spatial profiling over $\sim$1 $\mu$m$^2$ (see fig. \ref{fig:(a)-X-ray-diffraction}(c)) show very good uniformity with Mn : Pt ratio of 76 : 24 which results in a composition of Mn$_{3.04}$Pt$_{0.96}$. This slight deviation in stoichiometry with excess Mn concentration is not expected to destabilize the structure of Mn$_3$Pt, as reported in the study of hexagonal Mn$_3$Sn thin films \cite{Higo2018_Mn3Sn_Poly_AHE,Ji2006_MnPt_TF_Phasediagram}. Structural characterization of the films using X-ray diffraction (Rigaku diffractometer using Cobalt $K_{\alpha}$ source, $\lambda=$ 1.7903 \AA), performed in $\omega-2\theta$ geometry, show dominant (100) with weaker (200) peak and a (111) peak (see fig. \ref{fig:(a)-X-ray-diffraction}(d)). The indexed peaks are found to be in excellent agreement with the simulated pattern generated using VESTA software \cite{VESTA_Momma2011} by considering ideal lattice parameter for cubic structure of Mn$_3$Pt. The XRD peaks of 5 nm film are reduced in intensity and broader compared to the 25 nm film which indicates stronger lattice disorder. These results indicate polycrystalline growth of Mn$_3$Pt films with preferred out-of-plane orientation of the grains along [100] and [111] crystal axis. 

\section{Magnetic characterization}
\par
\begin{figure}
\includegraphics[width=8.5cm]{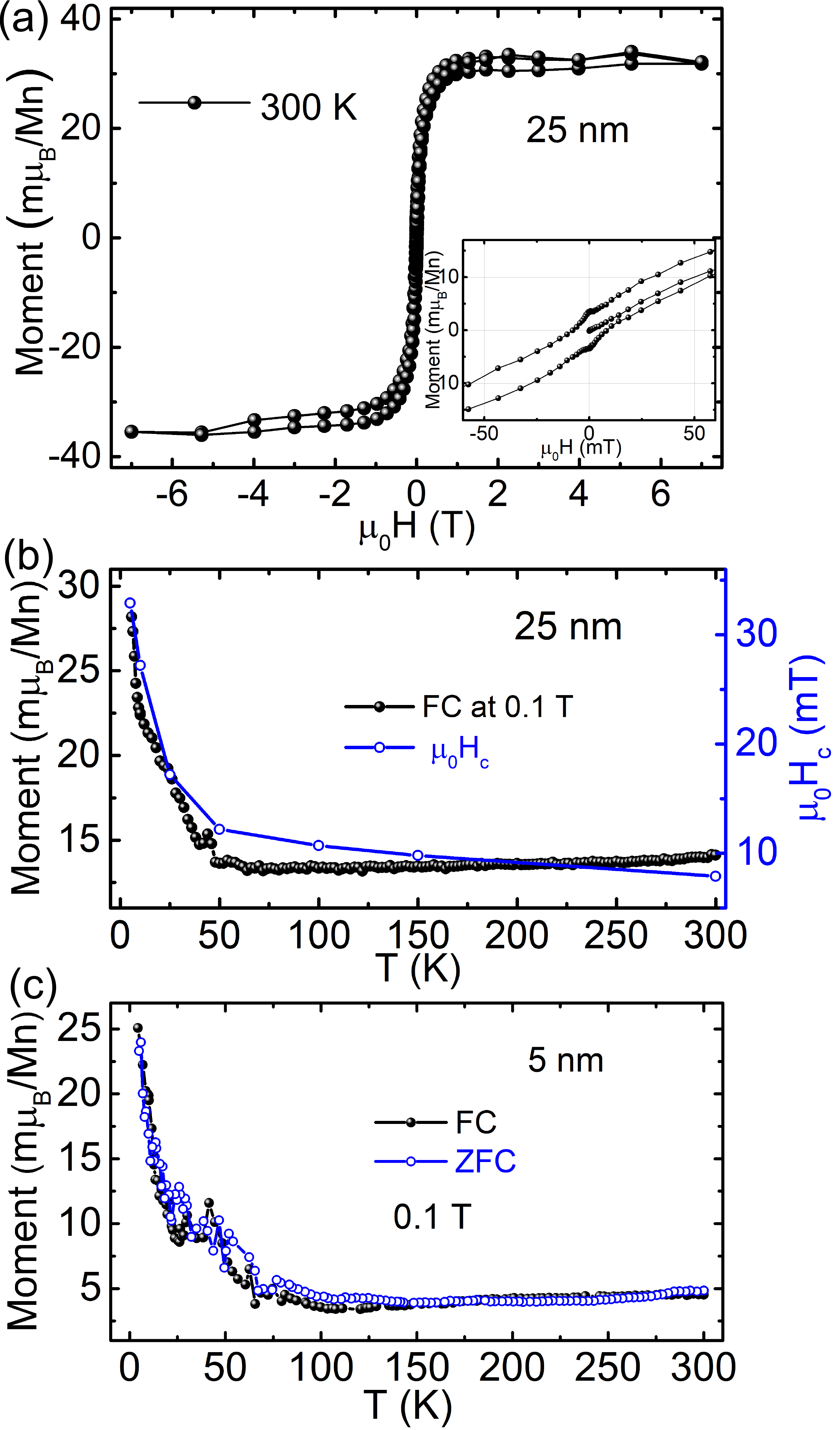}\caption{(a) M vs H curve of 25 nm Mn$_3$Pt thin film recorded at 300 K in OOP geometry. Inset shows zoomed-in view in the low-field range. (b) Temperature dependence of H$_c$ and M in 25 nm Mn$_3$Pt films under FC conditions. (c) M vs T of 5 nm Mn$_3$Pt thin film under FC and ZFC conditions. FC measurements were performed in an OOP magnetic field of 0.1 T. \label{fig:magnetic} }
\end{figure}

 The magnetic properties of Mn$_3$Pt films were investigated by recording the magnetization isotherms at different temperatures and by performing temperature dependent magnetization (M) measurements (see fig. \ref{fig:magnetic}). In these studies, the magnetic field (H) is applied normal to the film surface, referred henceforth as the out-of-plane (OOP) geometry. In figure \ref{fig:magnetic}(a), room-temperature M vs. H curve in the OOP geometry of the 25 nm films show a saturation magnetization of 32 m$\mu_{B}$/Mn with a coercivity (H$_c$) of 8 mT that is considerably smaller than the previous epitaxial work \cite{Liu2018_Mn3Pt_epitaxial_AHE}. This tiny magnitude of the magnetization rules out the formation of any ferromagnetic impurity phase in the Mn$_3$Pt thin films. Also, a smaller remanent magnetization of 3.2 m$\mu_{B}$/Mn is observed in these polycrystalline films which can appear when the applied magnetic field has a component along the kagome plane \cite{Nakatsuji2015_AHE_Mn3Sn}. This magnitude of the remanence is found to be similar to the previous reports in Mn$_3$Pt and Mn$_3$Sn samples \cite{Liu2018_Mn3Pt_epitaxial_AHE,Nakatsuji2015_AHE_Mn3Sn} supporting the formation of the NCAFM phase. Figure \ref{fig:magnetic}(b) shows the temperature dependence of H$_c$ of the above films to gradual increase from room temperature with a marked abrupt increase below $\sim$ 70 K reaching 33 mT at 5 K. Similar response is also observed in thinner films. Our field-cooled measurements (see fig. \ref{fig:magnetic}(b)) show a similar trend i.e. a small moment and temperature independent response followed by an abrupt increase below 70 K. Such a sharp rise in M and H$_c$ with temperature was previously reported in Mn$_3$Sn samples in the similar temperature regime \cite{Taylor2020_Mn3SnTF_AHE_THE,Rout2019_THE_Mn3Sn} that was associated with spin reorientation i.e. canting of Mn moments out of the kagome plane. However, unlike in the study of Mn$_3$Sn \cite{Rout2019_THE_Mn3Sn}, we do not observe any significant difference between the field-cooled and zero-field cooled measurements, as shown in figure \ref{fig:magnetic}(c) for the 5 nm films, indicating the absence of a spin-glass behaviour in the Mn$_3$Pt thin films.

\section{Magneto-transport measurements}
\par
\begin{figure}
\includegraphics[width=8.5cm]{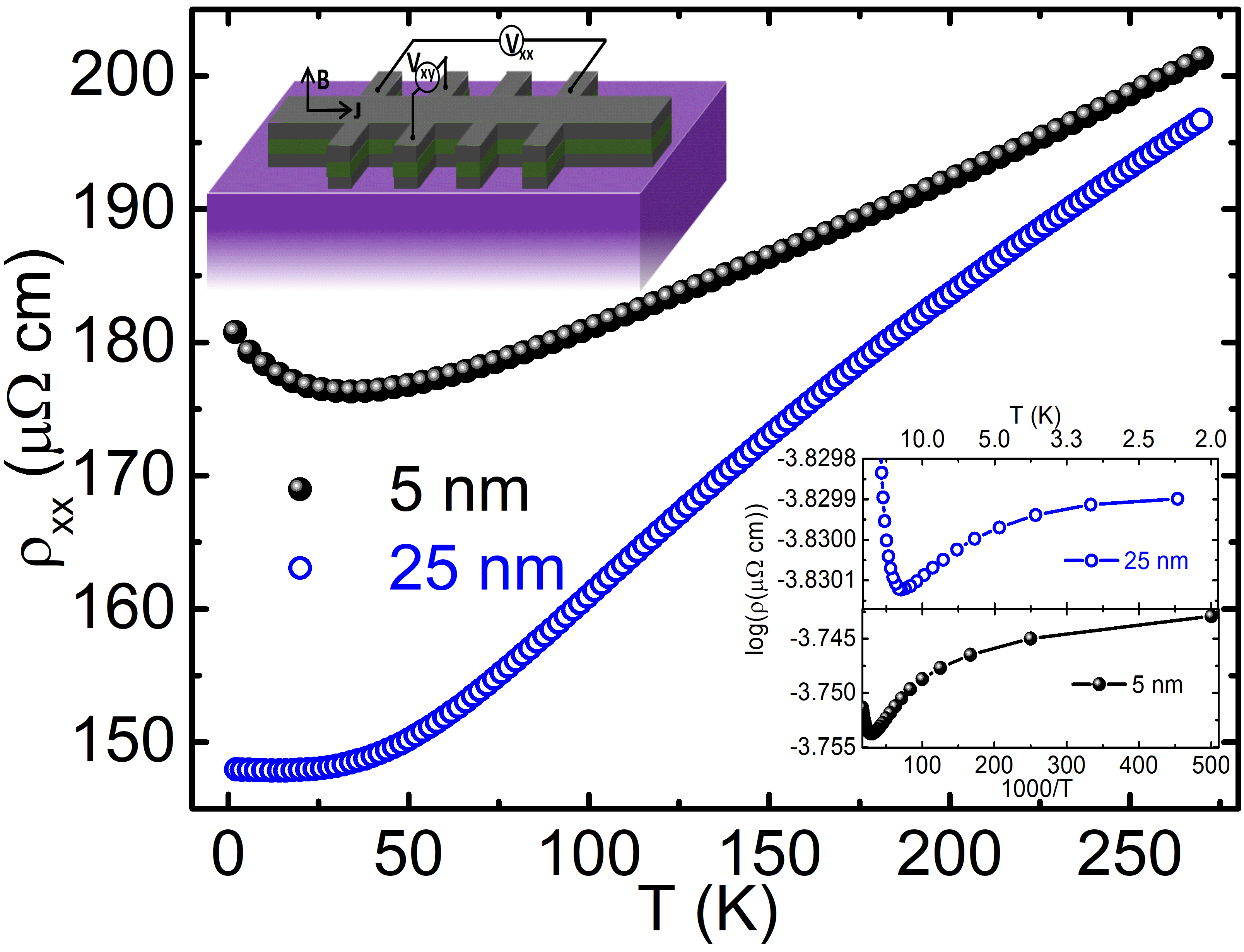}\caption{Temperature dependent longitudinal resistivity of 5 and 25 nm Mn$_3$Pt thin films grown on Si/SiO$_2$ substrates. Top inset shows the schematic of the Hall bar device. Bottom inset depicts $log\left(\rho\right)$ vs $1000/T$ plot of both thin films in the low temperature range.\label{fig:3_res}}
\end{figure}

 For the magnetotransport studies, Mn$_3$Pt films were patterned into a Hall bar geometry (length=1000 $\mu$m and width=100 $\mu$m) using a combination of optical lithography and argon ion milling (see top inset of Fig. \ref{fig:3_res}). At 277 K, longitudinal resistivity ($\rho_{xx}$) in our 25 and 5 nm films was found to be 195 and 205 $\mu\Omega$ cm, respectively, which is slightly higher than the Mn$_3$Pt epitaxial thin films \cite{Liu2018_Mn3Pt_epitaxial_AHE}. Here, $\rho_{xx}$ is obtained after subtracting the Ta contribution by considering a parallel circuit model \cite{Taylor2020_Mn3SnTF_AHE_THE}. In figure \ref{fig:3_res}, temperature variation of $\rho_{xx}$ showed a metallic response with a residual resistivity ratio ($RRR={\rho_{xx} (277 K)}/{\rho_{xx} (2 K)} $) of 1.36 and 1.15 in 25 and 5 nm devices, respectively. The smaller RRR in 5 nm film is expected due to the stronger lattice disorder compared to the thicker film. This signature is also revealed by a weak upturn below 12 K (35 K) in  25 nm (5 nm) films. The plot of $log\left(\rho\right)$ vs $1000/T$ in the low temperature regime (see fig. \ref{fig:3_res}(b)) do not show any linear dependence suggesting the absence of thermally activated band conduction, thereby attributing the above upturn to disorder effects \cite{Sugii2019_Res_Upturn_AHE_Mn3Sn,zhang2020many}. 

\begin{figure}
\includegraphics[width=8.5cm]{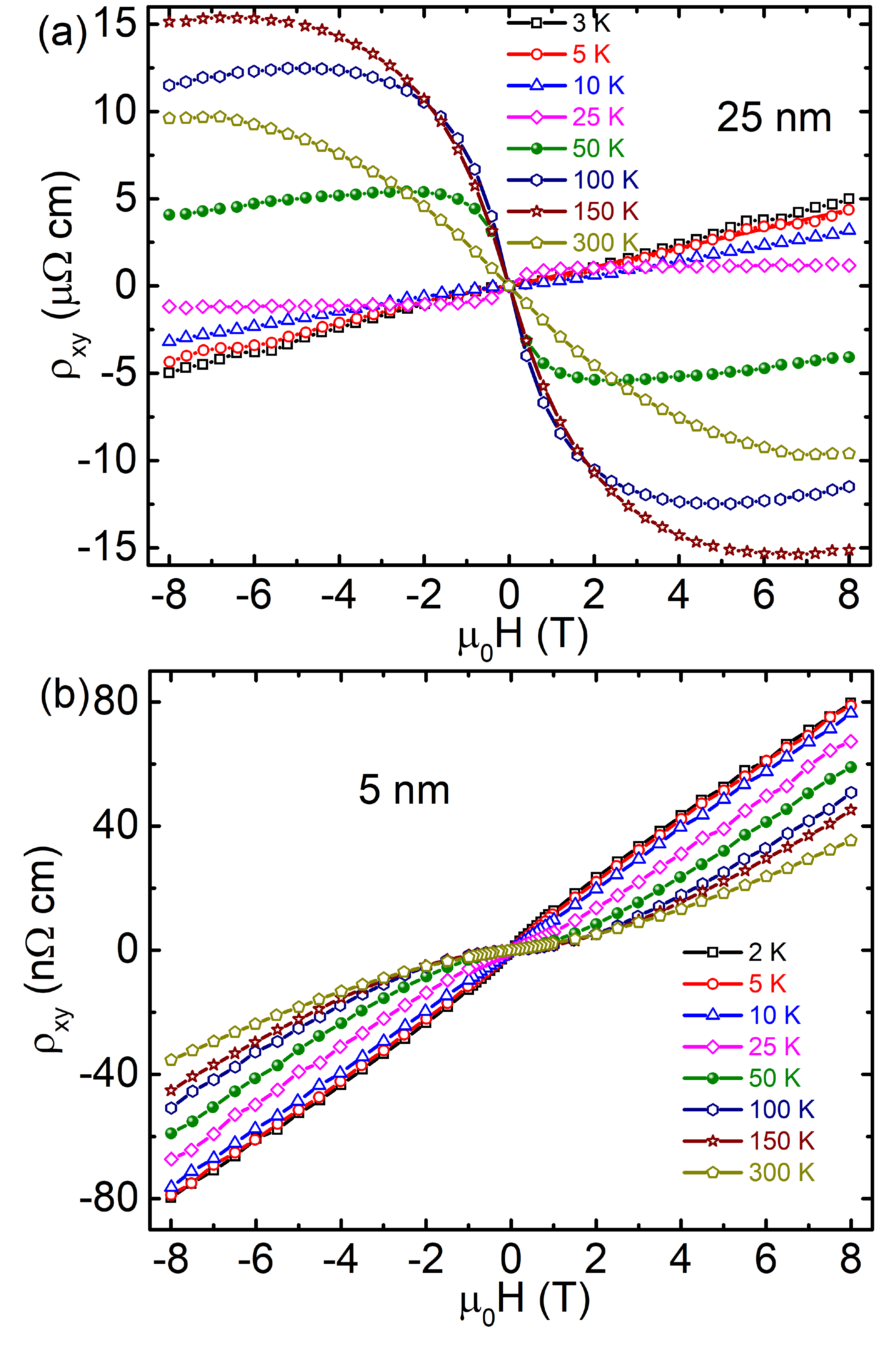}\caption{Temperature dependent transverse resistivity ($\rho_{xy}$) of (a) 25 nm and (b) 5 nm Mn$_3$Pt thin films grown on Si/SiO$_2$ substrate. Here, $\rho_{xy}$ is calculated by only considering  Mn$_3$Pt film thickness.\label{fig:4_vxy}}
\end{figure}

\begin{figure*}
\includegraphics[width=15cm]{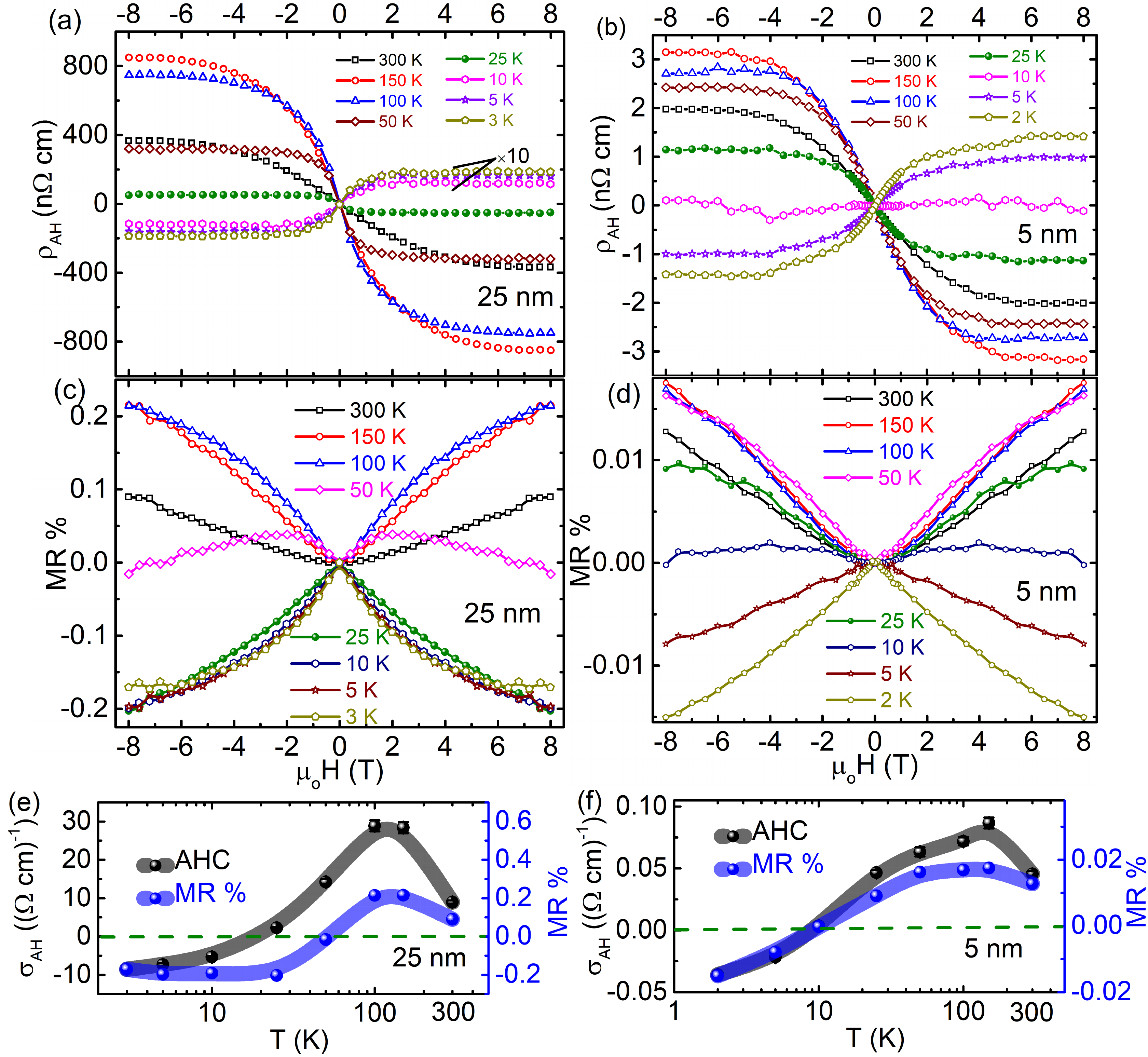}\caption{Temperature and magnetic field dependent anomalous Hall resistivity
of (a) 25 nm and (b) 5 nm Mn$_3$Pt films grown on Si/SiO$_2$
substrate. MR of (c) 25 nm and (d) 5 nm films measured at different temperature. Temperature variation of AHC and MR at 8 T in (e) 25 nm and (f) 5 nm films showing non-monotonic dependence with sign reversal at low temperature.
\label{fig_AHE_MR}}
\end{figure*}

The AHE of Mn$_3$Pt devices was investigated at different temperatures (300 K to 3 K) in the OOP geometry. Figures \ref{fig:4_vxy} (a) and (b) show the $\rho_{xy}$ vs H for the two extreme thickness limit of 25 nm and 5 nm. The total Hall resistivity is expressed by an empirical formula \cite{Pugh1930_AHE_eq.,Pugh1953_AHE_eq.},
\begin{equation}
\rho_{xy}=R_{0}\mu_{0}H+R_{S}M
\end{equation}
 where, $R_{0}$ and $R_{S}$ are the ordinary and anomalous Hall
coefficients respectively, and M is the magnetization. The first term of eq. (1) represents the Lorentz force induced ordinary Hall effect whereas the second term describes the anomalous Hall contribution. In both the devices, we observe nonlinear dependence of $\rho_{xy}$ with H. In 25 nm films, $\rho_{xy}$ is dominated by the nonlinear part indicating a major contribution of the anomalous Hall signal. Whereas, the 5 nm films show a large contribution of Lorentz force induced linear Hall signal. However, in both the devices, a positive slope of the linear Hall signal (at higher field) revealed p-type carriers suggesting the dominant contribution of hole-pockets at the Fermi surface (see Supplementary material for raw data) which is in-agreement with the band structure studies \cite{Liu2018_Mn3Pt_epitaxial_AHE}. The anomalous Hall resistivity ($\rho_{AH}$) was estimated by subtracting the linear part of the Hall voltage in the high magnetic field range (see fig. \ref{fig_AHE_MR} (a) and (b)). Here, $\rho_{AH}$ was found to be non-linear, saturating at high magnetic fields, with the curves transitioning from an inverted S-shaped to an S-shaped response at lower temperatures ($<$ $\sim$25 K). In these studies, we did not observe significant hysteresis in $\rho_{AH}$ at all the measured temperatures which is consistent with our magnetic studies showing a small coercive field. The above curves were also re-plotted in the scale of AHC, $\sigma_{AH}=-{\rho_{AH}}/{(\rho_{xx}^{2}+\rho_{xy}^{2})}$, demonstrating a similar trend. 

\par In figs. \ref{fig_AHE_MR} (c) and (d), we measured the magnetoresistance (MR $=$ ($(\rho_{xx}(H)-\rho_{xx}(0))/\rho_{xx}(0)\times100$)) in OOP geometry for the above two devices. We observe a non-saturating MR at high magnetic fields similar to the previous studies by Yan \emph{et al} \cite{Yan2019_TMR_Mn3Pt} in Mn$_3$Pt polycrystalline thin films. Interestingly, very similar non-monotonic temperature dependence of anomalous Hall signal and MR is revealed in figs. \ref{fig_AHE_MR} (e) and (f) for the measurements performed at 8 T. Here, the positive sign of AHC (or negative sign for $\rho_{AH}$) is consistent with the Berry curvature formalism reported in the work of Mn$_3$Sn \cite{Nakatsuji2015_AHE_Mn3Sn}. We find $\sigma_{AH}$ to initially increase from 9 $\Omega^{-1}$cm$^{-1}$ (0.05 $\Omega^{-1}$cm$^{-1}$) at 300 K to a maximum of 29 $\Omega^{-1}$cm$^{-1}$ (0.09 $\Omega^{-1}$cm$^{-1}$) around 100 K in the 25 nm (5 nm) Mn$_3$Pt devices. These values are roughly an order of magnitude smaller than the epitaxial Mn$_3$Pt film but are comparable with polycrystalline Mn$_3$Sn films  \cite{Higo2018_Mn3Sn_Poly_AHE}. The MR values also doubled when cooled from room temperature to 100 K.
\par At temperatures below 100 K,  we find MR and $\sigma_{AH}$ together drop in magnitude followed by a reversal in sign. MR and Hall voltage of bare Ta films, measured for reference, showed only positive MR with no anomalous Hall behavior, thereby ruling out the contribution of Ta in the above sign reversal. Such a feature of simultaneous sign reversal in MR and $\sigma_{AH}$ has not being reported previously in these systems.
We find our above observations in Mn$_3$Pt devices to indicate two competing effects. Comparing to our M vs T studies in figure \ref{fig:magnetic}, we associate the above device response to a transition from a Berry curvature dominated positive AHC effect to a trivial weak ferromagnetism related negative AHC effect at lower temperatures.  In the high temperature NCAFM phase, the spins in the three sublattices point at an angle of 120$^o$ relative to each other that are restricted within the kagome plane. Upon application of magnetic field, fluctuations in one of the sublattices, preferably the one with spins parallel to the field direction, may decrease while the other two may experience an increase in fluctuations leading to positive MR \cite{Yamada1973_MR_AFM,Nagasawa1972_Nd_MR_signchange,Hao2019_MR_fluctuation}. On the contrary, in a low temperature weak ferromagnetic phase caused by spin-reorientation or canting of Mn moments out-of-the-kagome plane, external fluctuations can be suppressed by the magnetic field giving rise to negative MR \cite{Yamada1973_MR_AFM}. Additionally, with a decrease in film thickness, the contribution of Berry curvature to AHC in the NCAFM phase with positive MR response weakens. This behaviour is consistent with our observation of decrease in the value of temperature at which we notice sign reversal of MR and AHC, occurring together at 10 K in 5 nm devices (see fig. \ref{fig_AHE_MR}(f)).

\section{Conclusion}
In summary, we have observed momentum space Berry-curvature induced room temperature AHE in NCAFM Mn$_3$Pt polycrystalline thin films with a dominant contribution from the hole carriers to the Hall response. Interestingly, we find a simultaneous sign reversal in AHC and MR values at lower temperatures. This sign reversal in both AHC and MR may be attributed to the spin reorientaion of Mn moments in Mn$_3$Pt from NCAFM to a canted ferromagnetic phase which is supported by the magnetic measurements. Temperature dependent spin-fluctuation studies using techniques such as  neutron reflectometry may provide a better understanding of the evolution of the magnetic structure of Mn$_3$Pt films at lower temperatures. Nonetheless, our present work has demonstrated the possibility to realise non-trivial topological states in the cubic phase of Mn$_3$Pt polycrystalline thin films at room temperature. The observations of large AHC and finite MR leads to the possibility of using Mn$_3$Pt as an interesting material system in the development of antiferromagnetic topological spintronics.

\section*{Acknowledgement}

\par J.M. and K.V.R acknowledge extra-mural funding from TIFR-Hyderabad and Department of Atomic Energy (DAE). We would like to thank Mr. Sasmal for fruitful discussions.



\bibliography{paper}
%

\onecolumngrid
\end{document}